\documentstyle[12pt,epsfig]{article}
\begin{document}
{\hskip 11.6cm} SNUTP 97-064\par
{\hskip 11.6cm} YUMS -97-13\par
{\hskip 11.6cm} hep-ph/9709386\par
{\hskip 11.6cm} November, 1997\\
\vspace{1ex}
\begin{center}        
{\LARGE \bf Four-quark Operators Relevant to B Meson Lifetimes 
from QCD Sum Rules}\\
\vspace{3ex}
{\sc M.S. Baek $^a$, Jungil Lee $^{b,}$\footnote{Address after Dec. 1, 1997:
Dept. of Phys., Ohio State Univ., Columbus OH43210, USA}, 
Chun Liu $^b$ and H.S. Song $^b$}\\
\vspace{2ex}   
{\it $^a$  Department of Physics, Yonsei University,}
{\it Seoul 120-749, Korea}\\  
{\it $^b$  Center For Theoretical Physics and Department of Physics}\\
{\it Seoul National University, Seoul 151-742, Korea}\\

\vspace{3.0ex}
{\large \bf Abstract}\\
\vspace{2ex}
\begin{minipage}{130mm}
                                                                               
   At the order of $1/m_b^3$, the B meson lifetimes are controlled by the 
hadronic matrix elements of some four-quark operators.  The nonfactorizable 
magnitudes of these four-quark operator matrix elements are analyzed by QCD 
sum rules in the framework of heavy quark effective theory.  The vacuum 
saturation for color-singlet four-quark operators is justified at hadronic 
scale, and the nonfactorizable effect is at a few percent level.  However for 
color-octet four-quark operators, the vacuum saturation is violated sizably 
that the nonfactorizable effect cannot be neglected for the B meson lifetimes.  
The implication to the 
extraction of some of the parameters from B decays is discussed.  The B meson 
lifetime ratio is predicted as $\tau(B^-)/\tau(B^0)=1.09\pm 0.02$.  However, 
the experimental result of the lifetime ratio $\tau(\Lambda_b)/\tau(B^0)$ 
still cannot be explained.
\par
\vspace{0.5cm}
\par
\end{minipage}
\end{center}

\newpage
                                                                               
   Heavy hadron lifetimes provide us with testing ground to the Standard 
Model, especially to QCD in some aspects [1-3], because they can be 
systematically calculated within the framework of heavy quark expansion. 
Theoretically, if we do not assume the failure of the local duality
assumption, the heavy hadron lifetime differences appear, at most, at the
order of $1/m_Q^2$ [4].  Recent experimental results on the lifetime ratio 
of $\Lambda_b$ baryon and B meson [5] showed some deviation from the 
theoretical expectation.  This has drawn a lot theoretical attentions [6-11].  
The current experimental values for the lifetime ratios which we are 
interested in are [5] 
\begin{equation}
\begin{array}{lll}
\displaystyle\frac{\tau(B^-)}{\tau(B^0)}&=&1.06\pm 0.04~,\\[3mm]
\displaystyle\frac{\tau(\Lambda_b)}{\tau(B^0)}&=&0.79\pm 0.06~.\\[3mm]
\end{array}
\end{equation}
This may imply that the $O(\frac{1}{m_b^2})$ contribution is not enough for 
the explanation of above heavy baryon and heavy meson lifetime difference.  
To the order of $1/m_b^3$, the hadron lifetimes have been studied since 
mid-80s [11, 12, 4, 6, 7].  And the potential importance of the $1/m_b^3$ 
corrections has been pointed out.   The parameterization of the hadronic 
matrix elements of four-quark operators which appear in the hadron lifetimes 
at the order of $1/m_b^3$ is generally expressed as [7]
\begin{equation}
\begin{array}{lcr}
\langle\bar{B}|\bar{b}\gamma_{\mu}(1-\gamma_5)q\bar{q}\gamma^{\mu}
(1-\gamma_5)b
|\bar{B}\rangle&\equiv&B_1F_B^2m_B^2~,\\
\langle\bar{B}|\bar{b}(1-\gamma_5)q\bar{q}(1+\gamma_5)b
|\bar{B}\rangle&\equiv&B_2F_B^2m_B^2~,\\
\langle\bar{B}|\bar{b}\gamma_{\mu}(1-\gamma_5)
t_aq\bar{q}\gamma^{\mu}(1-\gamma_5)
t_ab|\bar{B}\rangle&\equiv&\epsilon_1F_B^2m_B^2~,\\
\langle\bar{B}|\bar{b}(1-\gamma_5)t_aq\bar{q}(1+\gamma_5)t_ab
|\bar{B}\rangle&\equiv&\epsilon_2F_B^2m_B^2~,
\end{array}
\end{equation}
and 
\begin{equation}
\begin{array}{lcr}
\displaystyle\frac{1}{2m_{\Lambda_b}}
\langle\Lambda_b|\bar{b}\gamma_{\mu}(1-\gamma_5)q
\bar{q}\gamma^{\mu}(1-\gamma_5)b
|\Lambda_b\rangle&\equiv&\displaystyle-\frac{F_B^2m_B}{12}r~,\\[3mm]
\displaystyle\frac{1}{2m_{\Lambda_b}}
\langle\Lambda_b|\bar{b}(1-\gamma_5)q\bar{q}(1+\gamma_5)b|\Lambda_b\rangle
&\equiv&\displaystyle-\tilde{B}\frac{F_B^2m_B}{24}r~,\\[3mm]
\end{array}
\end{equation}
where the parameters $B_i$, 
$\epsilon_i$ ($i=1, 2$), $F_B$, 
$r$ and $\tilde{B}$ should be calculated by some 
nonperturbative QCD method.  In above equations, the renormalization scale is 
arbitrary, and the parameters depend on it.  It  
can be taken naturally at the low 
hadronic scale to apply heavy quark expansion.  On the other hand, if the 
scale is taken at $m_b$, parameter $F_B(m_b)$ is just the well-defined 
measurable physical quantity $-$ B meson decay constant $f_B$.  
\par
\vspace{1.0cm}    
   The QCD sum rule [13], which is regarded as a nonperturbative method
rooted in QCD itself, has been used successfully to calculate the properties 
of various hadrons.  In Ref. [8], in the framework of heavy quark effective 
theory (HQET), the baryonic parameters $r$ and $\tilde{B}$ have been 
calculated by QCD sum rule, $r\sim 0.1-0.3$, $\tilde{B}\simeq 1$.  For a
complete analysis, the mesonic parameters $B_i$ and $\epsilon_i$ should
be also calculated from QCD sum rule.  The four-quark operators, and hence 
$B_i$, $\epsilon_i$, $r$ and $\tilde{B}$, are scale-dependent quantities 
when the QCD radiative corrections are included.  It was proposed by Shifman 
and Voloshin [11] that at the low hadronic scale, the vacuum saturation 
approximation, namely $B_i=1$ and $\epsilon_i=0$, makes sense.  However in
this case, the measured lifetime ratio $\tau(\Lambda_b)/\tau(B^0)$ cannot be 
explained [8].  There are some argument, on the other hand, that the vacuum 
saturation maybe a poor approximation [9].  Especially from a naive large 
$N_c$ analysis, $\epsilon_i$'s are about $1/N_c\sim 0.3$ [11, 7].  We will 
explore the violation of the vacuum saturation approximation in detail from 
QCD sum rules in the framework of HQET.
\par
\vspace{1.0cm}
   Let us first consider the parameters $B_i$.  We construct the following 
three-point Green's function,
\begin{equation}
\Gamma^O(\omega, \omega')=i^2\int dxdye^{ik'\cdot x-ik\cdot y}
\langle0|{\cal T}[\bar{q}(x)\gamma^{\mu}\gamma_5h_v^{(b)}(x)]O(0)
[\bar{q}(y)\gamma_{\mu}
\gamma_5h_v^{(b)}(y)]^{\dag}|0\rangle~,
\end{equation}
where $\omega=2v\cdot k$, $\omega'=2v\cdot k'$; $h_v^{(b)}$ is the b-quark
field in the HQET with velocity $v$.  And $O$ denotes the color-singlet
operators given in Eq. (2), 
\begin{equation}
O=\bar{b}\Gamma_1 q\bar{q}\Gamma_2 b~,
\end{equation}
with $\Gamma_1=\Gamma_2=\gamma^{\mu}(1-\gamma_5)$ for $B_1$ and 
$\Gamma_1=1-\gamma_5$, $\Gamma_2=1+\gamma_5$ for $B_2$.  In terms of the
hadronic expression, the parameter $B_i$ appears in the ground state 
contribution of $\Gamma^O(\omega, \omega')$,
\begin{equation}
\Gamma^O(\omega, \omega')=B_i\frac{F_B^4m_B^2}
{(2\bar{\Lambda}-\omega)(2\bar{\Lambda}-\omega')}+ {\rm resonances}~,\\[3mm]
\end{equation}
where $\bar{\Lambda}=m_B-m_b$.  The 
resonance contribution will be simulated by the perturbative QCD contribution
above some threshold energy due to the local duality assumption.  On the 
other hand, this Green's function $\Gamma^O(\omega, \omega')$ will be 
calculated in terms of quarks and gluons, that is to say, by the operator 
product expansion method of QCD.  The essential feature of the QCD sum rule 
is that in the QCD calculation, the vacuum condensates of quarks and gluons 
have to be included.  Practically the calculation is performed at 1 GeV scale 
or so, only a few terms of the condensates with lowest dimensions are 
important.  Note that this calculation will be reinforced by the Borel 
transformation, and its consistency should be checked through the finding of 
the so-called sum rule window.  This procedure will be explained in more 
detail in the calculation of parameters $\epsilon_i$.
\par
\vspace{1.0cm}
   The calculation of $\Gamma^O(\omega, \omega')$ in HQET is straightforward.
The fixed point gauge [14] is adopted.  The assumption that the four-quark
condensates are factorizable is used.  The violation of it will be discussed 
later.  The dominant non-vanishing Feynman
diagrams are shown in Fig. 1 where the double lines denote the heavy quark.  
However, all these diagrams are factorizable, namely they do not contribute
any deviation from vacuum saturation.  And this is true even when the 
$O(\alpha_s)$ radiative corrections to these diagrams are included, simply 
because $O$ is a color-singlet operator.  Note that due to the same reason, 
there is no nonfactorizable gluon condensate contribution.  The tadpole 
diagrams in which the light quark lines from the four-quark vertex are 
contracted have been subtracted.  Generally, the 
nonfactorizable diagrams for a color-singlet four-quark operator are listed 
in Fig. 2.  They would be the leading contribution diagrams which might have 
an $O(1)$ correction to $B_i=1$.  But it is easy to see that they are 
vanishing because of the special structure of $\Gamma_1$ and $\Gamma_2$ 
in $O$ (Eq. (5)).  Hence, we see that the vacuum saturation approximation 
is valid at hadronic scale for color-singlet operators, $B_i=1$, for
$i=1,2$, through the leading order consideration.  
\par
\vspace{1.0cm}
   To what extent the nonfactorizable effect makes $B_i$ deviate from unity?
The next possibility for non-factorization is to consider two gluon 
exchanges, like those in Fig. 3.  It is interesting to note that the 
four-gluon condensate $<\alpha_s^2G^4>$ contribution to 
$\Gamma^O(\omega, \omega')$, like Fig. 3(b), is vanishing\footnote{This 
fact is gauge invariant and can be easily observed in the fixed point gauge.  
In this gauge, the light quark propagator with two condensate gluons attached
vanishes if one end of the propagator is at space-time origin [14].  And in 
this gauge, the heavy quarks in Figs. 1-3 are free from interaction.}.    
The perturbative two gluon 
exchange diagram, namely Fig. 3(a), can be the leading non-vanishing 
nonfactorizable contribution.  (There are some other diagrams which have the 
same or less order of magnitude numerically, like Fig. 3(c).)  This is a 
four-loop diagram, its contribution to $B_i$ is estimated as
$(\frac{\alpha_s(1 {\rm GeV})}{\pi})^2$ which is numerically around a few 
percent and is too small to be important for the hadron lifetimes.  We obtain 
that
\begin{equation}
B_i=1+O(10^{-2}),~~~~~(i=1~,~2)~.
\end{equation}
\par
\vspace{1.0cm}
   The parameters $\epsilon_i$ ($i$=1, 2), as we will see in the following 
analysis, have deviation from the expectation of the vacuum 
saturation at the hadronic scale which cannot be neglected for the B meson 
lifetimes.  The procedure is similar to that for 
parameters $B_i$.  The three-point Green's function in this case is 
constructed to be
\begin{equation}
\Gamma^T(\omega, \omega')=i^2\int dxdye^{ik'\cdot x-ik\cdot y}
\langle0|{\cal T}[\bar{q}(x)\gamma^{\mu}\gamma_5h_v^{(b)}(x)]
T(0)[\bar{q}(y)\gamma_{\mu} \gamma_5h_v^{(b)}(y)]^{\dag}|0\rangle~,
\end{equation}
where $T$ is the color-octet operators given in Eq. (2),
\begin{equation}
T=\bar{b}\Gamma_1 t_aq\bar{q}\Gamma_2 t_ab~.
\end{equation}
In the hadronic language, the parameter $\epsilon_i$ appears in the ground 
state contribution of $\Gamma^T(\omega, \omega')$,
\begin{equation}
\Gamma^T(\omega, \omega')=\epsilon_i\frac{F_B^4m_B^2}
{(2\bar{\Lambda}-\omega)(2\bar{\Lambda}-\omega')}+ {\rm resonances}~.\\[3mm]
\end{equation}
\par
\vspace{1.0cm}
   In the calculation of $\Gamma^T(\omega, \omega')$, all the condensates with 
dimensions lower than 6 are retained.  The dominant diagrams are
found to be those given in Fig. 4, and they are non-vanishing.  The four-quark 
condensate $\langle\bar{q}q\rangle^2$ diagram vanishes.  We have neglected 
the perturbative diagrams which are three-loop diagrams of order $\alpha_s$.  
That means the QCD radiative corrections to $\epsilon_i$ are not included.  
From general experience of QCD sum rule method, the condensate diagrams play 
more dominant role than the corresponding perturbative one.  This neglect is 
expected viable.  The condensates parameterize the non-perturbative effects
which in the 1 GeV scale are still small enough to be treated as power 
corrections of $1/\omega$ and $1/\omega'$ in the operator product expansion.
Because the perturbative contribution has been neglected, resonances in 
Eq. (10) will be also neglected due to the duality assumption.  While the 
calculation can be justified if ($-\omega$) and ($-\omega'$) are large, 
however the hadron ground state property should be obtained at small 
($-\omega$) and ($-\omega'$).  These contradictory requirements can be 
achieved by introducing double Borel transformation for $\omega$ and 
$\omega'$.  There are two Borel parameters $\tilde{T}$ and $\tilde{T'}$ 
corresponding to $\omega$ and $\omega'$, respectively.  They appear 
symmetrically, so we take $\tilde{T} = \tilde{T'}$ in the analysis.  
\par
\vspace{1.0cm}
   The sum rules for the parameters $\epsilon_i$ are 
\begin{equation}
\begin{array}{lll}
\displaystyle\epsilon_1&=&\displaystyle
m_0^2\langle\bar{q}q\rangle \frac{\tilde{T}}{32\pi^2}
\frac{e^{4\bar{\Lambda}/\tilde{T}}}{F_B^4m_B^2}~,\\[3mm]
\displaystyle\epsilon_2&=&\displaystyle
-\left(\frac{\langle\alpha_sGG\rangle}{4\pi}\tilde{T}
+m_0^2\langle\bar{q}q\rangle\right) \frac{\tilde{T}}{16\pi^2}
\frac{e^{4\bar{\Lambda}/\tilde{T}}}{F_B^4m_B^2}~,\\[3mm]
\end{array}
\end{equation}
There is no gluon 
condensate contribution to $\epsilon_1$.  
Numerically we use the following 
values of the condensates,
\begin{equation}
\begin{array}{lll}
\langle\bar{q}q\rangle&\simeq&-(0.23~ {\rm GeV})^3~,\\
\langle\alpha_sGG\rangle&\simeq&0.04~ {\rm GeV}^4~,\\
\langle g\bar{q}\sigma_{\mu\nu}G^{\mu\nu}q\rangle&\equiv&
m_0^2\langle\bar{q}q\rangle~,
~~~~~m_0^2\simeq0.8~{\rm GeV}^2~.\\[3mm]
\end{array}
\end{equation}
For consistence, the HQET sum rule of the parameter $F_B$ [15, 16] will 
be used.  The result derived in Ref. [16] is
\begin{equation}
F_B^2m_Be^{-2\bar{\Lambda}/\tilde{T}}=\frac{3}{8\pi^2}\int_0^{\omega_c}d\nu
\nu^2e^{-\nu/\tilde{T}}-
\langle\bar{q}q\rangle
\left(1-\frac{m_0^2}{4\tilde{T}^2}\right)~,\\[3mm]
\end{equation}
where $\omega_c$ is twice of the continuum threshold, which was determined as
$\omega_c\simeq 2.0\pm0.3$ GeV.  The range of the Borel parameter $\tilde{T}$ 
in Eq. (11) should be similar to that of $F_B$.  This point can be obviously 
seen from the sum rules for $B_i$ if it is written down explicitly.  In that 
sum rules, if all the non-factorizable contributions are neglected, then the 
resulting $B_i$ is unity only if the Borel parameter is the same as that of 
$F_B$.  Practically we take the window as $0.7\leq\tilde{T}\leq 1.0$ GeV.  
There is no $\bar{\Lambda}$ dependences for $\epsilon_i$'s, because they are 
canceled actually.  The numerical sum rule results for $\epsilon_1$ and 
$\epsilon_2$ are given in Fig. 5.  From the figures, we see that 
$\epsilon_i$'s have no good stability in the window of $\tilde{T}$.  
This is because we have not included the perturbative diagram.  Finally the 
results for $\epsilon_1$ and $\epsilon_2$ are obtained as 
\begin{equation}
\begin{array}{lll}
\epsilon_1&\simeq&-(4.1\pm2.2)\times 10^{-2}~,\\[3mm]
\epsilon_2&\simeq&(6.1\pm3.5)\times 10^{-2}~.\\[3mm]
\end{array}
\end{equation}
In spite of the large uncertainties in above equation, the numbers are 
significant for the B meson lifetime difference.  We see that the vacuum 
saturation approximation for the color-octet matrix elements is indeed 
violated.  The magnitudes of $\epsilon_i$ can be as large as 0.1. 
\par
\vspace{1.0cm}
   It is necessary to discuss the hypothesis on the four-quark condensate
factorization.  We have used it in above analysis for $B_i$ and $\epsilon_i$.  
While this is the working assumption for the usual QCD sum rule calculations, 
the violation of it may imply in our case, new contribution to the 
non-factorizable effect of the four-quark operator matrix elements.  This 
contribution to $B_i$ and $\epsilon_i$ is estimated as 
\begin{equation}
\delta B_i\simeq \delta \epsilon_i =\delta\langle\bar{q}q\rangle^2 
\frac{e^{4\bar{\Lambda}/\tilde{T}}}{F_B^4m_B^2}~,
\end{equation}
where $\delta\langle\bar{q}q\rangle^2$ denotes the deviation of the four-quark
condensates from factorization.  Arguments based on large $N_c$ expansion 
suggest that this approximation is good to within $1/N_c^2\sim 10\%$ [17].  
Numerically, considering the success of this assumption in the calculation of 
baryons by QCD sum rules [18], we take $\delta\langle\bar{q}q\rangle^2$ to be 
less than $30\%$ of $\langle\bar{q}q\rangle^2$.  In this case, $\delta B_i$ 
and $\delta \epsilon_i$ are smaller than about $10^{-2}$ which
have no significant influence on our results for $B_i$ and $\epsilon_i$.
\par
\vspace{1.0cm}
   It should be noted that the vacuum saturation and its violation we have 
analyzed above are at some hadronic scale, other than the scale of $m_b$.  
Because we have been working in the framework of HQET, in which the natural 
scale is $\mu_{had}\ll m_b$.  The renormalization group evolution of the
relevant operators was calculated in Refs. [19, 7].  Information on 
parameters $B_i$ and $\epsilon_i$ at scale $m_b$ is necessary to obtain the 
hadron lifetimes by using the analysis of Ref. [7].  There is a notation 
difference in this paper.  In HQET, $F_B$ given in Eq. (13) is in fact a 
scale-dependent quantity if the renormalization effect is considered.  From 
this point of view, $F_B$ in our previous analysis took value at hadronic 
scale, whereas that in Ref. [7] at scale of $m_b$.  Taking this difference 
into consideration, we have the following evolution relations by choosing
$\alpha_s(\mu_{had})= 0.5$ (corresponding to $\mu_{had}\sim 0.67$ GeV),
\begin{equation}
\begin{array}{lll}
B_i(m_b)& \simeq& B_i(\mu_{had})-0.24\epsilon_i(\mu_{had})~,\\[3mm]
\epsilon_i(m_b)& \simeq& -0.05B_i(\mu_{had})+0.72\epsilon_i(\mu_{had})~.
\end{array}
\end{equation}
That is 
\begin{equation}
\begin{array}{llllll}
B_1(m_b)& \simeq& 1.01\pm 0.01~,& ~~~~~
B_2(m_b)& \simeq& 0.99\pm 0.01~,\\[3mm]
\epsilon_1(m_b)& \simeq& -0.08\pm 0.02~,& ~~~~~
\epsilon_2(m_b)& \simeq& -0.01\pm 0.03~,
\end{array}
\end{equation}
from Eq. (14).
\par
\vspace{1.0cm}
   To be more relevant, we discuss a more detailed parameterization for the
four-quark operators proposed in Ref. [6] in the following.  It is motivated
for the model-independent determination of the hadronic matrix elements 
from the lepton spectrum in the endpoint region of semileptonic B decays.
The parameterization is written as
\begin{equation}
\begin{array}{lcr}
\langle\bar{B}|\bar{b}\gamma_{\mu}(1-\gamma_5)q\bar{q}\gamma_{\nu}
(1-\gamma_5)b|\bar{B}\rangle&\equiv&(v_{singl}v_{\mu}v_{\nu}-g_{singl}
g_{\mu\nu})F_B^2m_B^2~,\\
\langle\bar{B}|\bar{b}\gamma_{\mu}(1-\gamma_5)t_aq\bar{q}\gamma_{\nu}
(1-\gamma_5)t_ab|\bar{B}\rangle&\equiv&(v_{oct}v_{\mu}v_{\nu}-g_{oct}
g_{\mu\nu})F_B^2m_B^2~,\\
\end{array}
\end{equation}
where $v_{singl}$, $g_{singl}$ and $v_{oct}$, $g_{oct}$ are the parameters.
The tadpole diagram
contributions of the matrix elements have been subtracted.  The parameters
are related to those of Eq. (2) as $v_{singl}-4g_{singl}=B_1$ and
$v_{oct}-4g_{oct}=\epsilon_1$.  In the vacuum saturation approximation, 
$v_{singl}=1$, $g_{singl}=v_{oct}=g_{oct}=0$.  They can be calculated by QCD
sum rules.  From our above analysis for the color-singlet operator, we have
\begin{equation}
v_{singl}\simeq 1+O(10^{-2}), ~~~~~g_{singl}\simeq O(10^{-2}).
\end{equation}
The sum rules for the color-octet operator are obtained as
\begin{equation}
\begin{array}{lll}
\displaystyle v_{oct}&=&\displaystyle
-\left(\frac{\langle\alpha_sGG\rangle}{6\pi}\tilde{T}+m_0^2
\langle\bar{q}q\rangle\right)\frac{\tilde{T}}{8\pi^2}
\frac{e^{4\bar{\Lambda}/\tilde{T}}}{F_B^4m_B^2}~,\\[3mm]
\displaystyle g_{oct}&=&\displaystyle
-\left(\frac{\langle\alpha_sGG\rangle}{12\pi}\tilde{T}+m_0^2
\langle\bar{q}q\rangle\right)\frac{\tilde{T}}{16\pi^2}
\frac{e^{4\bar{\Lambda}/\tilde{T}}}{F_B^4m_B^2}~.\\[3mm]
\end{array}
\end{equation}
The numerical results for $v_{oct}$ and $g_{oct}$ are given in Fig. 6 from 
which we obtain
\begin{equation}
v_{oct}\simeq (1.35\pm 0.76)\times 10^{-1}~, ~~~~~
g_{oct}\simeq  (0.74\pm 0.40)\times 10^{-1}~.
\end{equation}
They are typically at the order of 0.1.
\par
\vspace{1.0cm}
   From the QCD sum rule point of view, the vacuum saturation for the           
hadronic matrix elements of the color-singlet four-quark operators given in
Eq. (5) or in the first equation of (18) is valid up to a few percent level.  
However, that for the color-octet four-quark operators is violated at ten
percent level.
The result that the nonfactorizable effect for color-octet operators is more
significant can be also understood qualitatively, if we look at the 
perturbative diagrams.  This effect is at three-loop level, whereas that for 
color-singlet operators at the four-loop level.
\par
\vspace{1.0cm}
   The $1/m_b$ corrections to above results can be analyzed in principle.
While having little influence on our above arguments and calculations, they
formally belong to the $O(1/m_b^4)$ effects to the b-hadron lifetimes.
\par
\vspace{1.0cm}
   In Ref. [6], the extraction of the parameters from B decays are discussed.
The role of parameter $g_{singl}$ might be significant in comparing the 
decay rates of $B^0\rightarrow l\nu X_u$ and $B^-\rightarrow l\nu X_u$.  
From our analysis, the difference of the decay rates should be small due to
$g_{singl}$ is very small even at the scale of $m_b$.  The ratio of
$g_{singl}/g_{oct}$ can be obtained from the lepton spectrum of the above 
decays.  While having not determined the sign of $g_{singl}$, our analysis
prefers an even smaller ratio of $|g_{singl}/g_{oct}|\sim 1/10$ than Ref. [6]
did.  These points will be verified by the experiments in the near future.
\par
\vspace{1.0cm}
   The b-hadron lifetime ratios are expressed as [7]
\begin{equation}
\begin{array}{lll}
\displaystyle
\frac{\tau(B^-)}{\tau(B^0)}&=&1+0.03B_1-0.71\epsilon_1+0.20\epsilon_2
+O(\frac{1}{m_b^4})~,\\
\displaystyle
\frac{\tau(\Lambda_b)}{\tau(B^0)}&=&0.98-0.17\epsilon_1+0.20\epsilon_2
-(0.013+0.022\tilde{B})r+O(\frac{1}{m_b^4})~.
\end{array}
\end{equation}
In above equation, the energy scale for the parameters is at $m_b$.  the B 
meson decay constant $f_B$
is taken as 200 MeV.  By using the results in Eq. (17), we obtain 
\begin{equation}
\frac{\tau(B^-)}{\tau(B^0)}\simeq 1.09\pm 0.02~~~~~~~{\rm and}~~~~~~~
\frac{\tau(\Lambda_b)}{\tau(B^0)}\simeq 0.98\pm 0.01~.\\[3mm]
\end{equation}
In obtaining above numbers, the QCD sum rule results for the baryonic 
parameters $r$ and $\tilde{B}$ [8] have been also taken values at scale 
$m_b$.  From Eq. (23) we see that, while $\tau(B^-)/ \tau(B^0)$ is agree 
with measurement, the QCD sum rule result for $\tau(\Lambda_b)/ \tau(B^0)$,
after including $1/m_b^3$ corrections, still contradicts the experimental
measurement. 
\par
\vspace{1.0cm}
   In summary, the nonfactorizable contributions of the hadronic matrix 
elements of four-quark operators relevant to the B meson lifetimes have been 
studied by QCD sum rules in the framework of HQET.  The vacuum saturation 
for color-singlet four-quark operators is justified at hadronic scale, and 
the nonfactorizable effect is at a few percent level.  However, the vacuum 
saturation for color-octet four-quark operators is violated sizably that the 
nonfactorizable effect cannot be neglected for the B meson lifetimes.  
The implication to the 
extraction of some of the parameters from B decays has been discussed.  The 
B meson lifetime ratio has been predicted to be consistent with experiment,
$\tau(B^-)/\tau(B^0)=1.09\pm 0.02$.  However, the discrepancy between theory 
and experiment on $\tau(\Lambda_b)/ \tau(B^0)$ has not been solved.  It is 
unlikely that higher $1/m_b$ corrections will give the solution.  If the 
experimental data of $\tau(\Lambda_b)/\tau(B^0)$ is further confirmed in the 
future, that may imply either the failure of local duality assumption or some 
new physics.  
\par
\vspace{2.0cm}
\begin{center}
{\bf Acknowledgments}
\end{center}
   We are grateful to Matthias Neubert for important comments and 
S.Y. Choi and Pyungwon Ko for helpful discussions.  C.L. would like to thank
Yasuhiro Okada and KEK Theory Group for hospitality where part of the work
was finished.  This work was supported in part by KOSEF through the SRC and
MOE of Korea through the BSR Program (BSRI-97-2418).  J.L. was supported by 
KOSEF Fellowship.  MSB was supported by KOSEF and RUF of Yonsei University 
through MOE.

\newpage

{\Large \bf Figure captions}\\

Fig. 1.  Dominant non-vanishing Feynman diagrams for 
$\Gamma^O(\omega, \omega')$.

Fig. 2.  Nonfactorizable diagrams for a general color-singlet 
four-quark operator.

Fig. 3.  Nonfactorizable diagrams by two gluon exchanges for 
$\Gamma^O(\omega, \omega')$.

Fig. 4.  Condensate contribution to $\Gamma^T(\omega, \omega')$.

Fig. 5.  Sum rules for $\epsilon_1$ (a) and $\epsilon_2$ (b).
The sum rule window is $\tilde{T}=0.7-1.0$ GeV.

Fig. 6.  Sum rules for $v_{oct}$ (a) and $g_{oct}$ (b).
\newpage
{\Large \bf Figures}\\
\vskip 2cm
\begin{center}
\epsfig{file=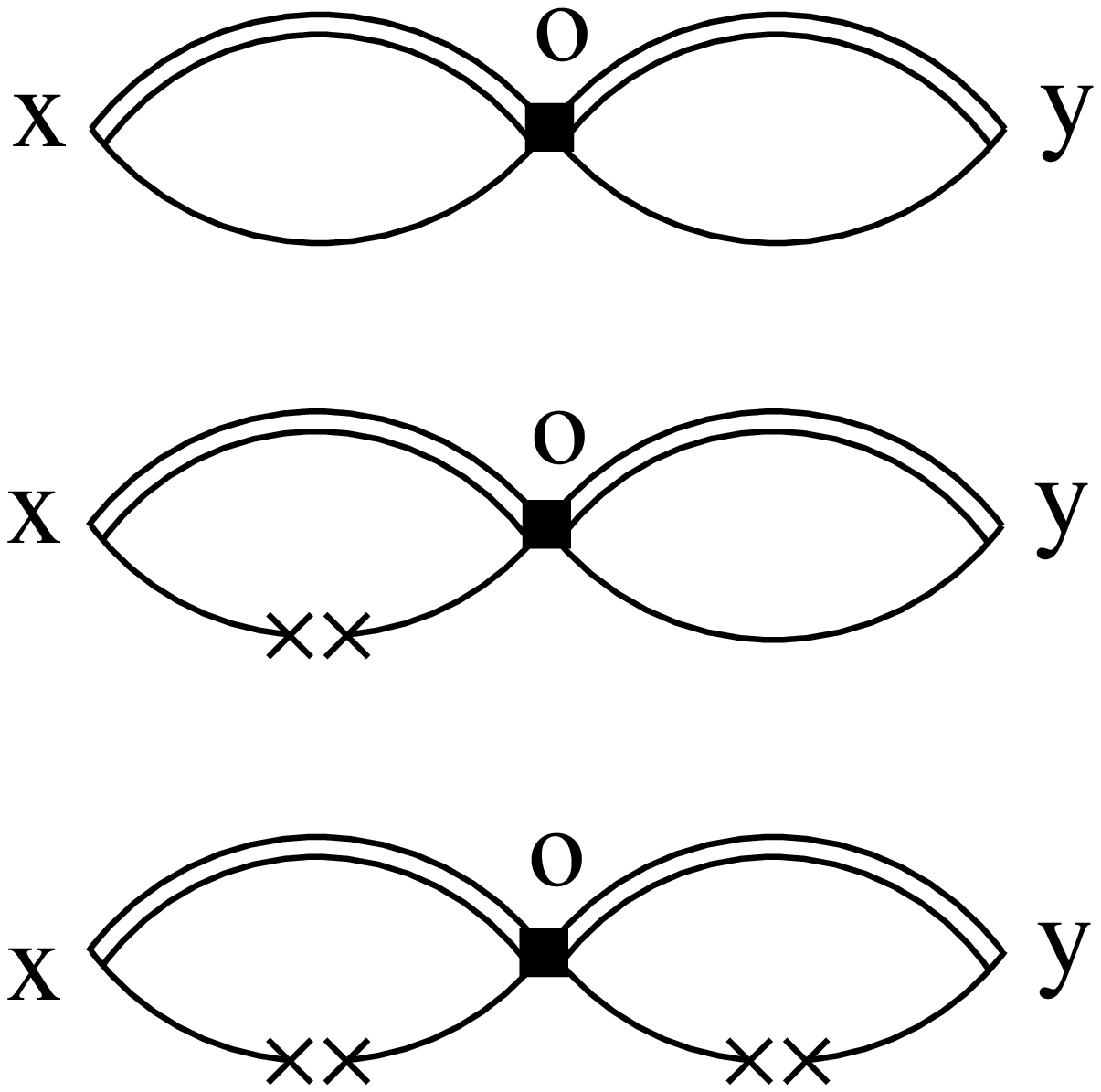,height=12cm}
\vskip.5cm
{\large\bf Fig.~1}
\end{center}
\begin{center}
\epsfig{file=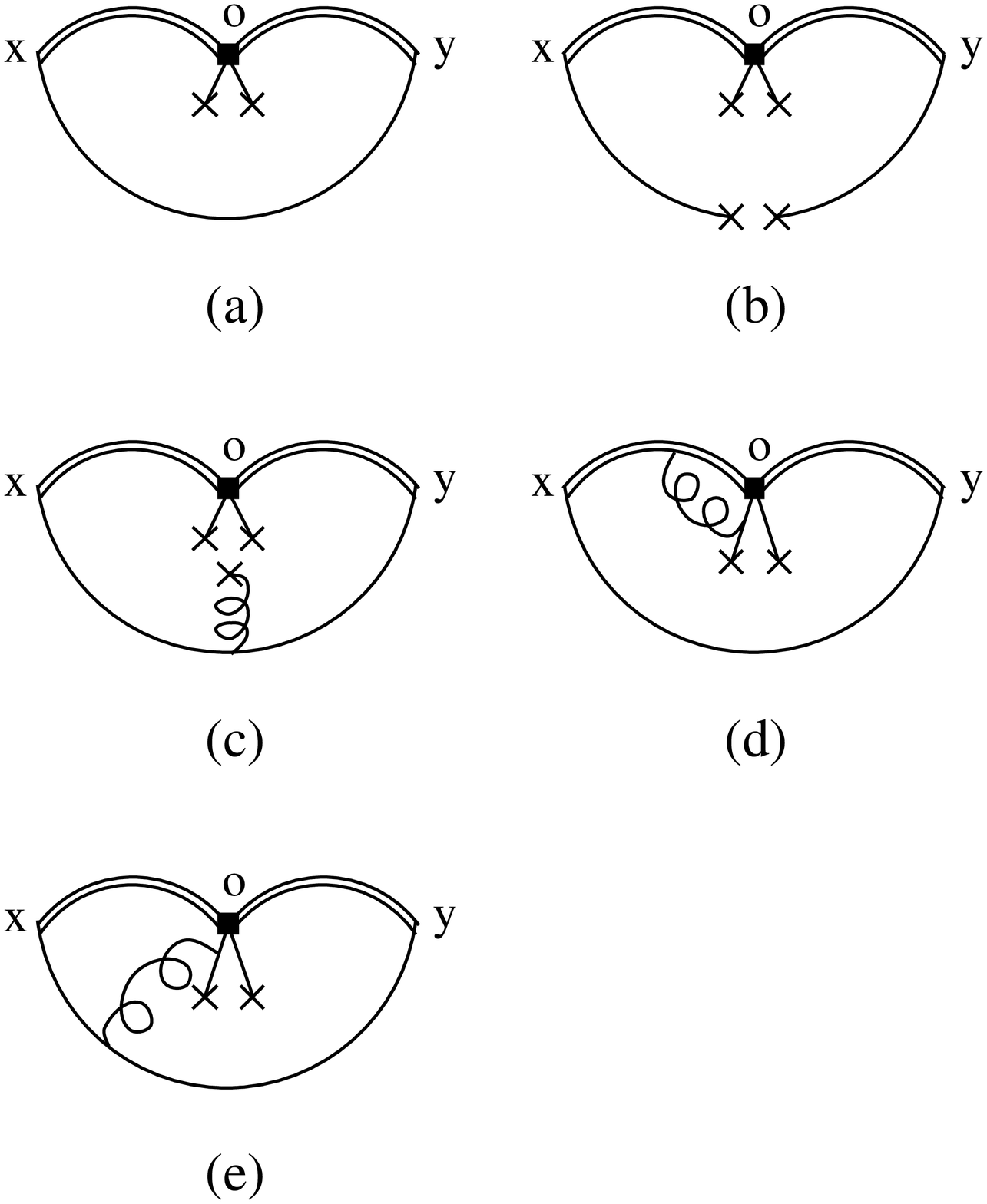,height=16cm}
\vskip.5cm
{\large\bf Fig.~2}
\end{center}
\begin{center}
\epsfig{file=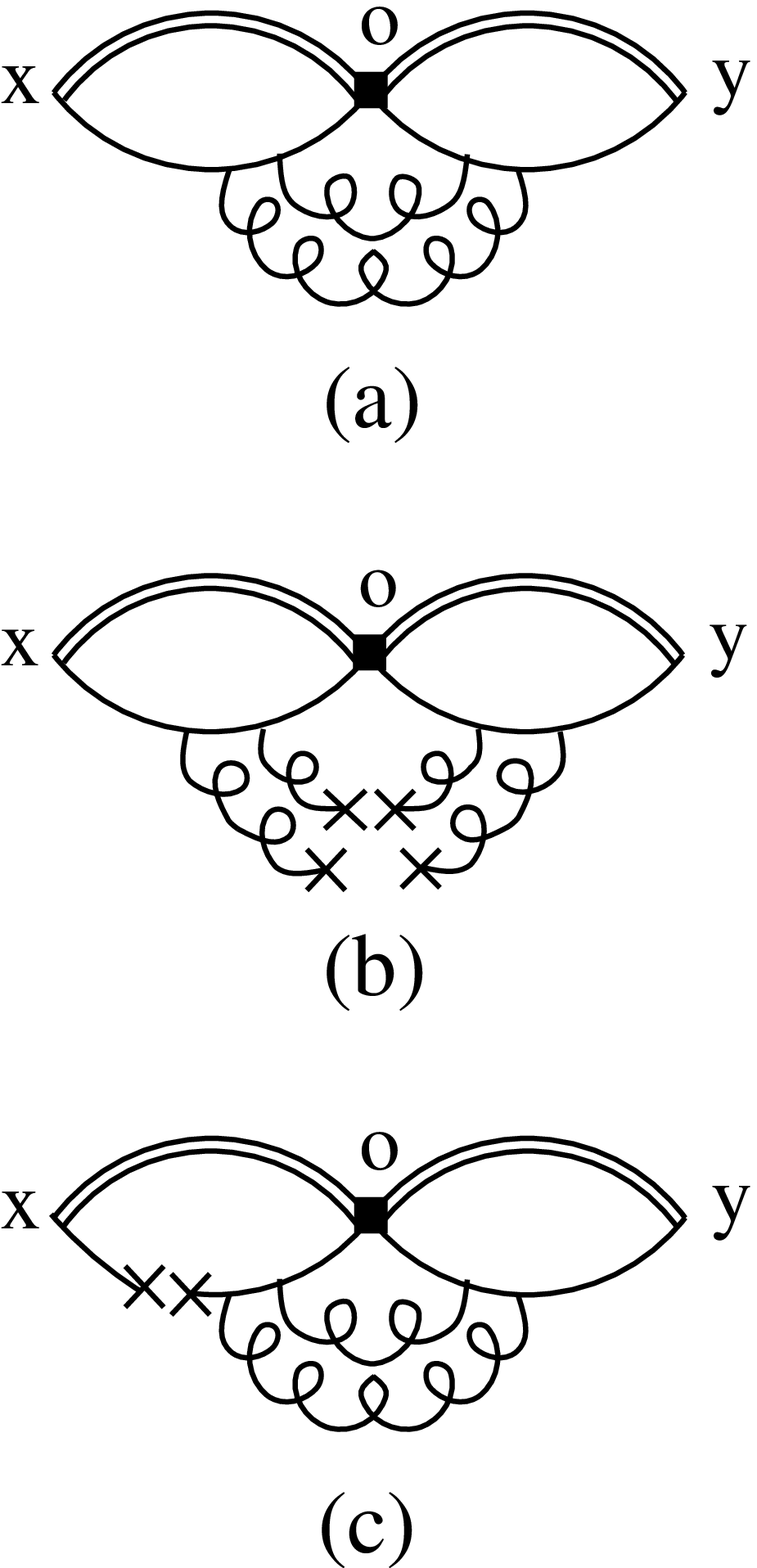,height=16cm}
\vskip.5cm
{\large\bf Fig.~3}
\end{center}
\newpage
\begin{center}
\epsfig{file=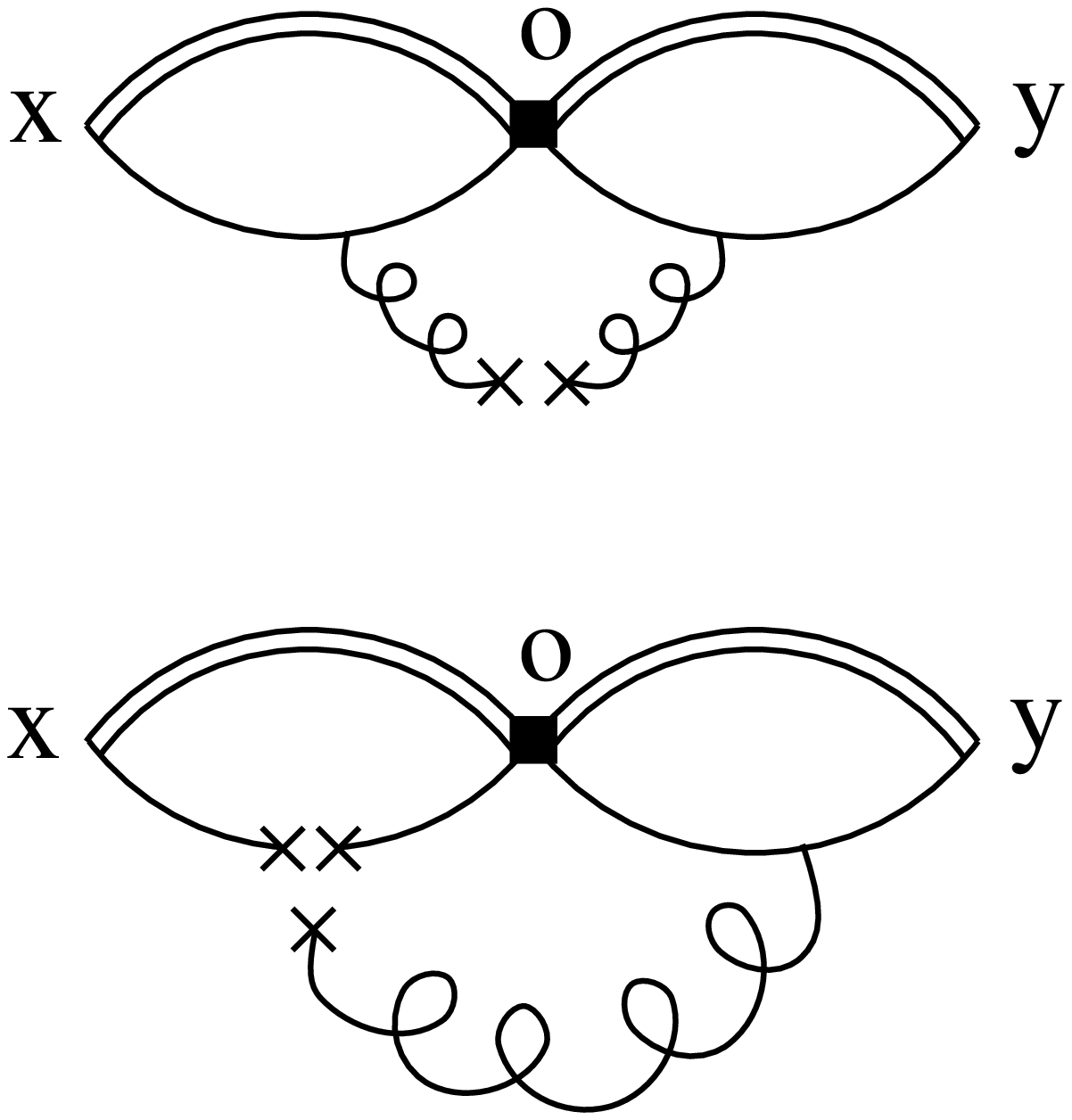,height=8.5cm}
\vskip.5cm
{\large\bf Fig.~4}
\end{center}
\newpage
\begin{center}
\epsfig{file=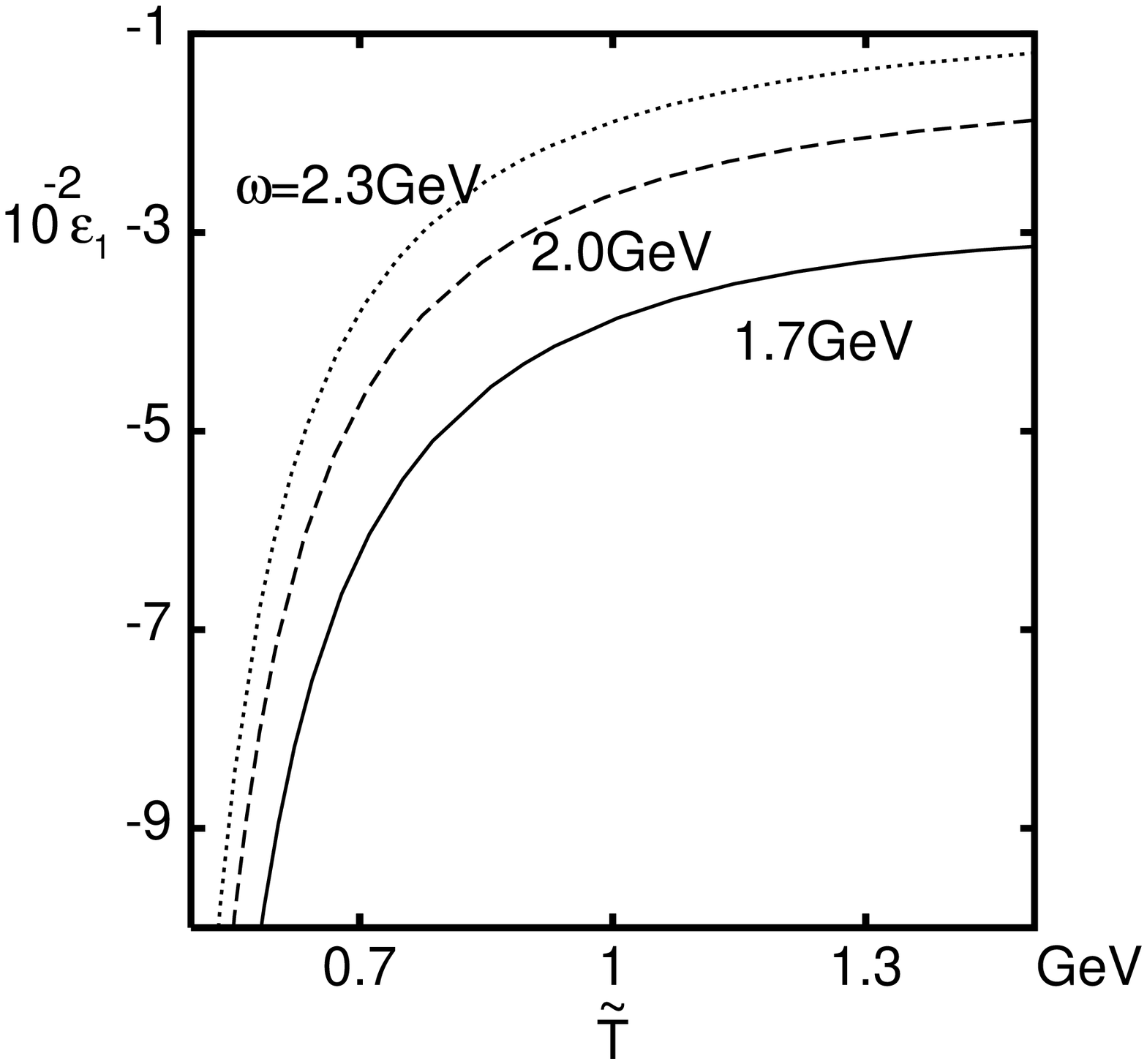,height=8.5cm}
\vskip.5cm
{\large\bf Fig.~5(a)}
\end{center}
\vskip1cm
\begin{center}
\epsfig{file=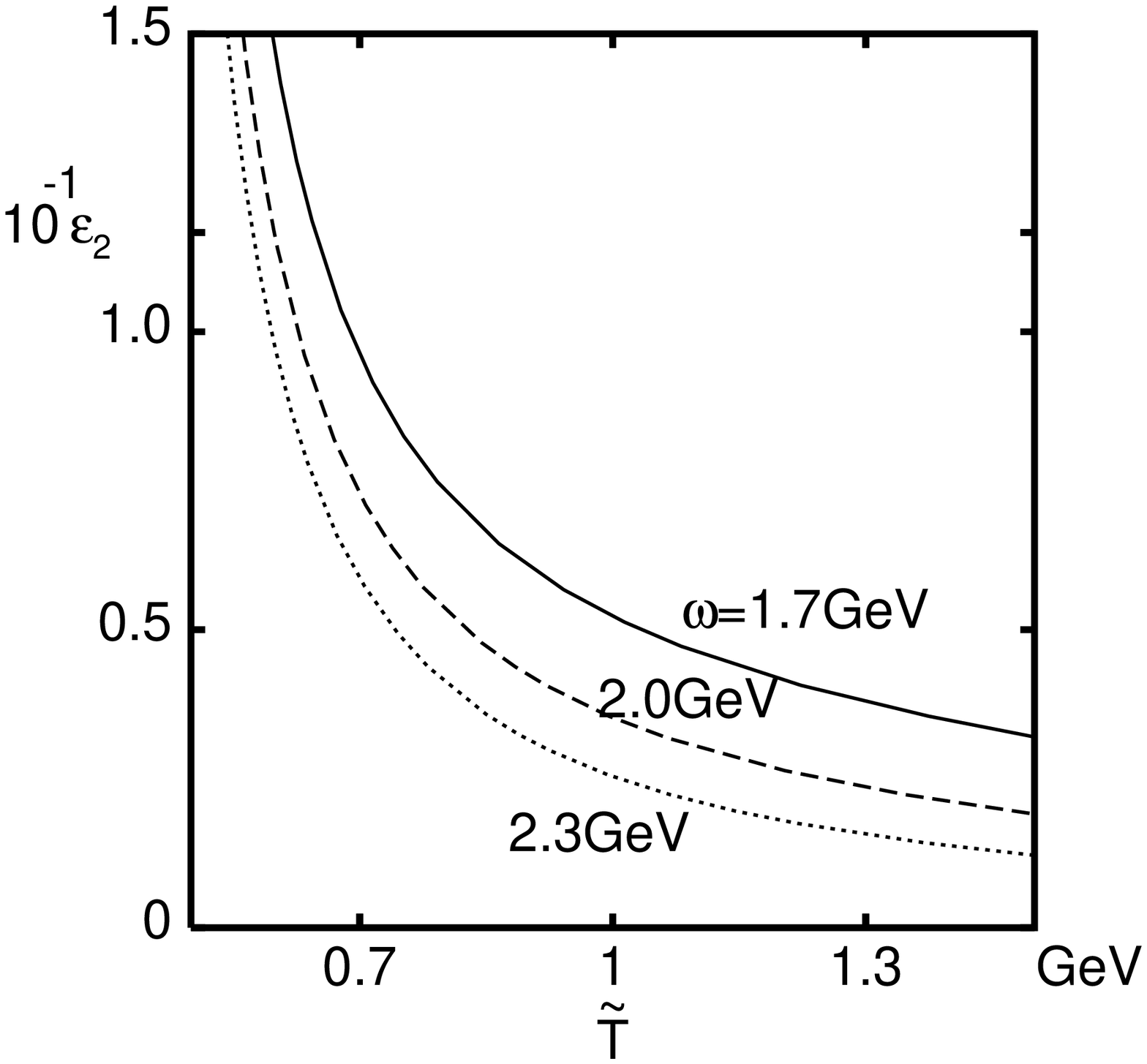,height=8.5cm}
\vskip.5cm
{\large\bf Fig.~5(b)}
\end{center}
\newpage
\begin{center}
\epsfig{file=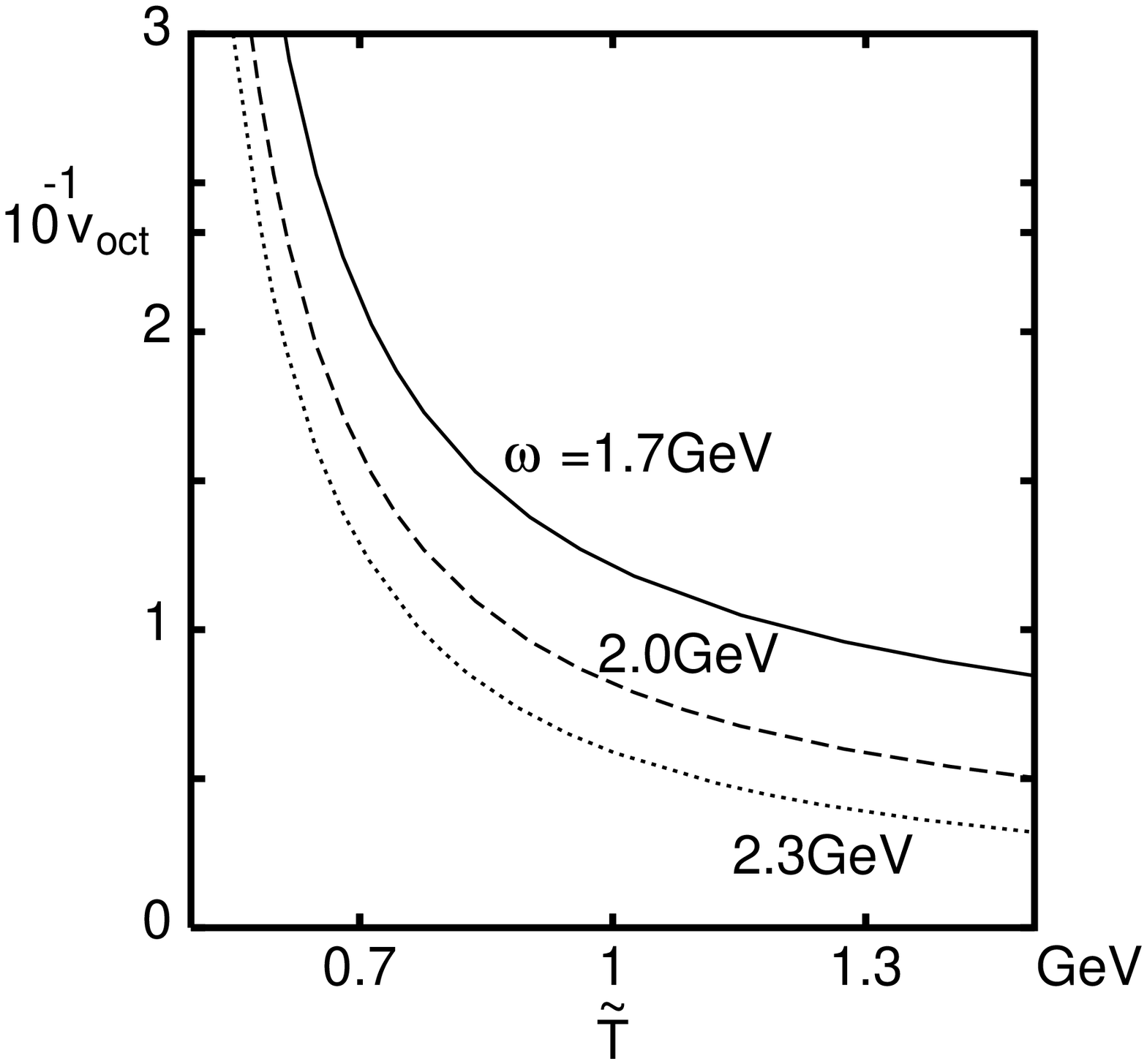,height=8.5cm}
\vskip.5cm
{\large\bf Fig.~6(a)}
\end{center}
\vskip1cm
\begin{center}
\epsfig{file=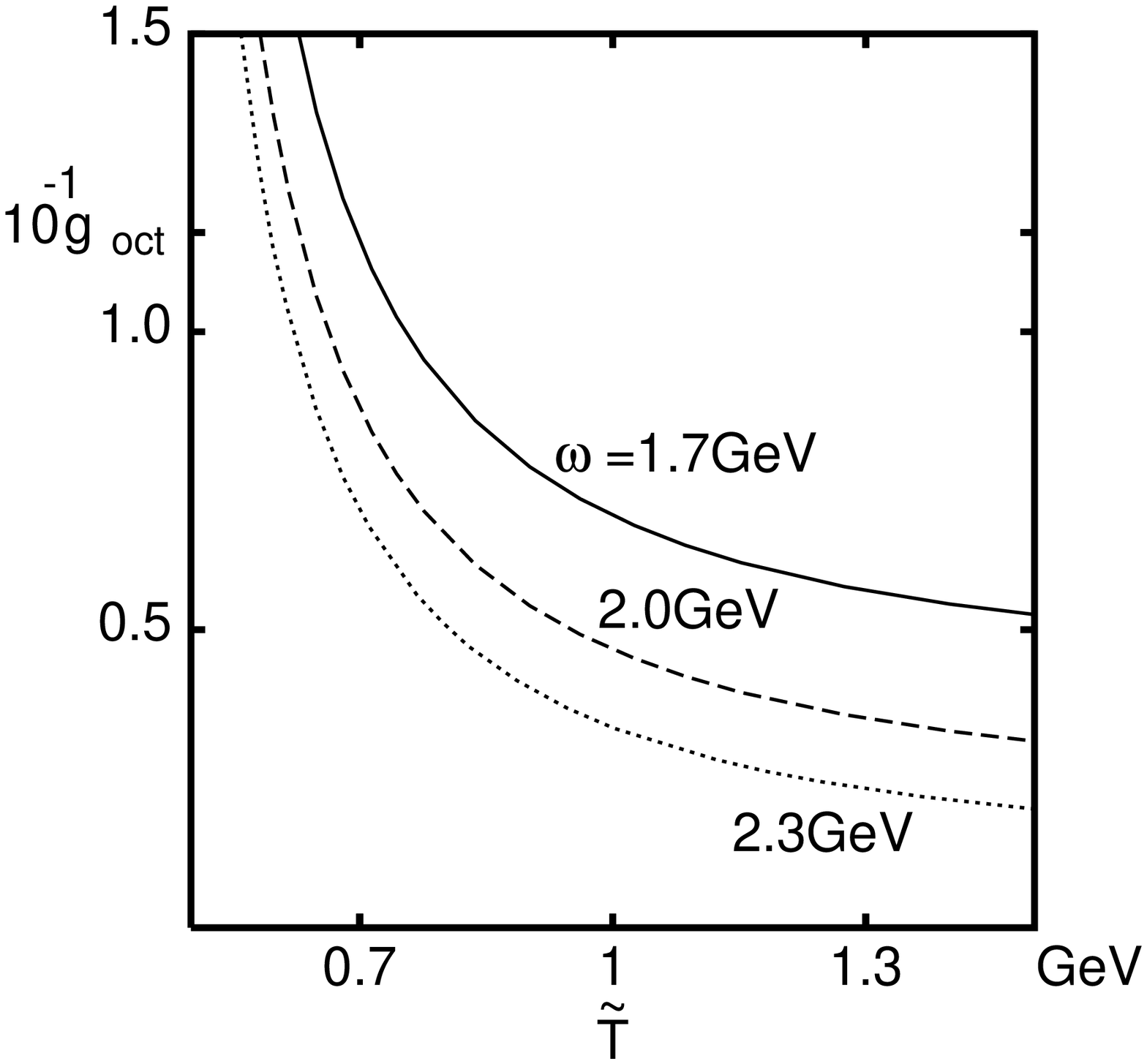,height=8.5cm}
\vskip.5cm
{\large\bf Fig.~6(b)}
\end{center}

\end{document}